\documentclass[english]{emulateapj}

\usepackage[T1]{fontenc}
\usepackage[latin1]{inputenc}
\usepackage{graphicx}
\usepackage{amssymb}

\providecommand{\tabularnewline}{\\}

\makeatletter

\makeatletter

\usepackage{times}

\shorttitle{The Triple Origin of Blue Stragglers}
\shortauthors{Perets and Fabrycky}

\makeatother

\makeatother

\usepackage{babel}

\makeatother

\usepackage{babel}

\usepackage{babel}

\begin{document}

\title{On the triple origin of blue stragglers}

\author{Hagai B. Perets\altaffilmark{1} \& Daniel C. Fabrycky\altaffilmark{2,3}}

\altaffiltext{1}{Weizmann Institute of Science, Rehovot 76100,
Israel; \texttt{hagai.perets@weizmann.ac.}}

\altaffiltext{2}{Harvard-Smithsonian Center for Astrophysics,
60 Garden St, MS-51, Cambridge, MA 02138}

\altaffiltext{3}{Michelson Fellow} 
\begin{abstract}
Blue straggler stars (BSSs) are stars observed to be hotter and bluer
than other stars with the same luminosity in their environment. As
such they appear to be much younger than the rest of the stellar population.
Two main channels have been suggested to produce such stars: (1) collisions
between stars in clusters or (2) mass transfer between, or merger
of, the components of primordial short-period binaries. Here we suggest
a third scenario, in which the progenitor of BSSs are formed in primordial
(or dynamically formed) hierarchical triple stars. In such configurations
the dynamical evolution of the triples through the Kozai mechanism
and tidal friction can induce the formation of very close inner binaries.
Angular momentum loss in a magnetized wind or stellar evolution could
then lead to the merger of these binaries (or to mass transfer between
them) and produce BSSs in binary (or triple) systems. We study this
mechanism and its implications and show that it could naturally explain
many of the characteristics of the BSS population in clusters, most
notably the large binary fraction of long period BSS binaries; their
unique period-eccentricity distribution (with typical periods$>$700
days); and the typical location of these BSSs in the color-magnitude
diagram, far from the cluster turn-off point of their host clusters.
We suggest that this scenario has a major (possibly dominant) role
in the formation of BSSs in open clusters and give specific predictions
for the the BSSs population formed in this manner. We also note that
triple systems may be the progenitors of the brightest planetary nebulae
in old elliptical galaxies, which possibly evolved from BSSs. 
\end{abstract}

\section{Introduction}

Blue Straggler Stars (BSSs) are stars that appear to be anomalously
young compared to other stars of their population. In particular,
BSSs lie along an extension of the main sequence (MS) in the color-magnitude
diagram, a region from which most of the stars of equal mass and age
have already evolved. Such stars appear to be brighter and bluer than
the turn-off point of the stellar population in which they were observed.
Their location in the color magnitude diagram suggests that BSSs have
typical masses of $1.2-1.5\, M_{\odot}$, that are significantly larger
than those of normal stars in old stellar systems such as old open
clusters (OCs) or globular clusters (GCs). Thus, they are thought
to have increased their mass during their evolution. Two main mechanisms
have been proposed for their formation: $i)$ the merger of two stars
induced by stellar collision \citep{hil+76} and $ii)$ coalescence
or mass-transfer between two companions in a binary system \citep{mcc64}.
The roles of each of these mechanisms in producing the observed BSSs
populations are still debated, as each of these scenarios were found
to be successful in explaining some of the BSSs observations, but
fail in others (e.g., \citealt{bai95}). In fact, even when both these
mechanisms are taken into account (e.g., in N-body simulations including
stellar evolution), they have major difficulties explaining the observations,
especially those of binary BSSs \citep{Leo96,hur+05}: the period-eccentricity
distribution of BSSs binaries produced through these mechanisms are
in poor agreement with the observed distribution of BSS binaries (\S\ref{sec:pedist}).
Moreover, typical BSS binaries produced in combined N-body and stellar
evolution simulations are produced in the inner regions of a cluster
core \citep{hur+05}, whereas observations show many of the BSS binaries
in clusters to exist much farther out \citep{gel+08,gel+09}.

In the binary merger scenario for BSS formation, close binaries with
periods shorter than $5-6$ days evolve into mass transfer configuration
or merger in less than $10$ Gyrs, thus producing a rejuvenated star
\citep{and+06}. As we point out in this work theoretical studies
and observations suggest that most such close binaries form as the
inner binaries in triple systems \citep{kis+98,2006EKE,2006T,2007FT}.
A straight-forward conclusion is that BSSs formed in close binaries
are most likely to be (or to have been) members of triple systems.
Such a scenario has strong implications on the properties of BSSs,
their multiplicity and their orbital parameters. Here we raise this
basic conclusion and follow its implications. We suggest a third mechanism
for the origin of BSSs, in which the progenitors of blue stragglers
are formed in primordial (and also in dynamically-formed) hierarchical
triple stars. The inner binary in such triples can be rapidly driven
into close or even contact configurations, due to the combined effects
of Kozai cycles and tidal friction (KCTF mechanism: \citealp{kis+98,2001EK,2007FT}),
as we discuss below. Such close binaries could then evolve through
mass transfer or merger and produce BSSs. We show that such a scenario
could explain and predict the characteristics of the BSS population
in clusters, and could naturally explain the large fraction of long
period binary BSSs, their unique period-eccentricity distribution
and their location in the color magnitude diagram, whereas previously
proposed mechanisms cannot. This mechanism has some additional predictions
that could discriminate it from other models for BSSs formation: (1)
BSSs could have long-period \emph{main sequence} binary companions,
(2) the spin axis of such BSSs are likely to be misaligned in a specific
way from the orbital axis of the binary orbit, and (3) they may exist
in regions where collisions between stars are unlikely; additional
implications are discussed below. Such predictions differ from those
of the previously suggested mechanisms for BSSs formation and could
serve to further support and confirm the novel model suggested here.

We note that although triples have been suggested before to play a
role in BSSs formation, they were not considered to have a major contribution
to the BSSs population \citep{Leo96,iva08}, and the role of primordial
triples has not been discussed. Here we study the triple origin scenario
for BSSs and its general implications in detail. We show that current
observations strongly support this model and confirm its predictions
where other models fail.

In this paper we begin with an overview of the dynamics of Kozai cycles
and tidal friction in triples (\S\ref{s:overview}). Then we compare
the timescales for the KCTF mechanism and for binary disruption in
different environments (\S\ref{cluster BSS}) and describe the theoretical
and observational implications of this formation mechanism for various
environments (\S\ref{sub:BSSs-low density}). We then summarize in
\S \ref{s:Summary}.

\section{Triple dynamics, Kozai cycles and tidal friction}

\label{s:overview}

To be stable for many orbital periods, triple systems usually require
a hierarchical configuration in which two stars orbit each other in
a relatively tight {}``inner binary'', and the third star and the
inner binary orbit their common center of mass as a wider {}``outer
binary''. Although such triples do not disrupt, their orbits may
change shape and orientation on timescales much longer than the dynamical
time. A particularly important change was discovered by \citet{1962K},
who studied the orbital changes of asteroids due to the weak interactions
with Jupiter (where the asteroid-Sun system serves as an inner binary
in the Sun-Jupiter-asteroid triple). He found that if the asteroid's
initial inclination relative to Jupiter's orbit is high enough, secular
torques will cause its eccentricity and inclination to fluctuate out
of phase with one another: these are called {}``Kozai oscillations''.
\citet{1962L} independently studied a similar process effecting the
motion of artificial satellites of the Earth as they are perturbed
by the Sun and Moon, and he noted the possibility of collision between
a satellite and the Earth if the satellite's eccentricity becomes
large enough.

Collisions were prominent also in the first application of these dynamical
concepts to triple stars. \citet{1968H} noted that large initial
inclination ($i_{c}\lesssim i\lesssim180^{\circ}-i_{c}$, for a {}``Kozai
critical angle'' of $i_{c}\approx40^{\circ}$) leads to large eccentricities,
which could cause a tidal interaction, mass loss, or even collision
of the members of the inner binary. Thus, Harrington reasoned that
a triple star system with an inner binary mutually perpendicular to
the outer binary should not exist for many secular timescales. However,
as was noted by \citet{1979MS}, the inner binary stars coming close
to one another will not merge immediately; instead, the tidal dissipation
between them shortens the semi-major axis of the inner binary during
these eccentricity cycles. They suggested that such inner binaries
could therefore attain a very close configuration, in which mass transfer
and accretion could occur, possibly forming cataclysmic variables
or binary X-ray sources. \citet{2006EKE} discussed binary stellar
evolution, including mergers, following KCTF. Recently, \citet{iva08}
discussed the possibility of forming BSSs in dynamically formed triples
in dense clusters, showing that such newly formed triples may explain
as much as $10\%$ of the BSSs in such clusters.

The equations of motion from the equilibrium tide model, with arbitrary
eccentricity of a binary system and arbitrary spin obliquities of
its components, were coupled to triple star dynamics by \citet{1998EKH}.
This analysis led to further fruitful studies \citep{kis+98,2001EK},
suggesting that a large percentage of close binaries may have become
close through KCTF. Several observational studies verified that close
binaries very often have tertiary components \citep{2004T,2006PR,2006DKR,2007RPV},
showing a strong dependence between the binary period and the existence
of a third companion \citep{2006T}. In fact it was found that nearly
all ($\sim96\%$) closest binaries ($P<3$ days) have distant tertiary
components \citep{2006T}. Recently \citet{2007FT} used the equations
of \citet{1998EKH} to verify that the observational results of \citet{2006T}
are consistent with KCTF acting on a population of triples. They found
that the population of close binaries could be explained through evolution
in triples, even if no such primordial binaries (with $P<6$ days)
exist. They also noticed that although eccentricity will have damped
during the tightening of the close binary, the mutual inclination
between inner and outer binaries should very often finish at either
$i\approx40^{\circ}$ or $i\approx140^{\circ}$.

\begin{figure}
\begin{tabular}{c}
\includegraphics[clip,width=0.95\columnwidth]{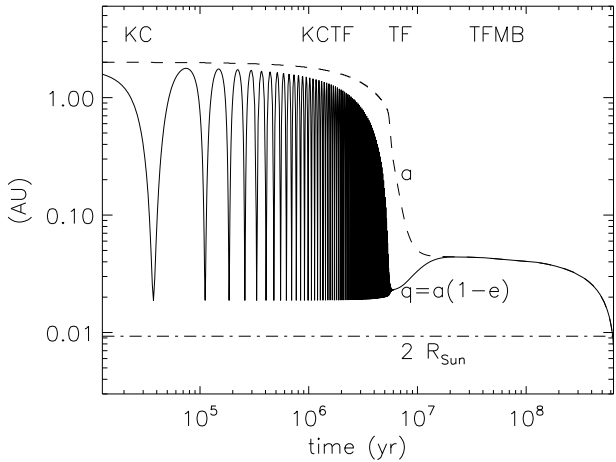}\tabularnewline
\end{tabular}

\caption{\label{fig:kctfmb}Merger of the two stars of an inner binary, accomplished
by a combination of Kozai cycles (KC), tidal friction (TF), and magnetic
braking (MB). }

\end{figure}

The connection to BSS comes when the inner binary merges to become
a single, more massive star. We put the whole scenario together in
Figure~\ref{fig:kctfmb}, assuming the mechanism for close binary
merger is angular momentum loss through magnetized stellar winds.
The initial system is an inner binary of two solar mass stars, at
low eccentricity and $a=2$~AU, orbited by a 0.5 $M_{\odot}$ star
on a circular orbit at $50$~AU with mutual inclination of $84^{\circ}$.
On short timescales, the eccentricity of the inner binary fluctuates
(Kozai cycles; KC). On millions of year timescales, tidal friction
seals in a large eccentricity (KCTF), then damps the binary at constant
orbital angular momentum (TF). Lastly, on a timescale of $\sim1$~Gyr,
magnetic braking (MB) of the stellar spins drains the orbital angular
momentum because the spins stay tidally locked, causing the binary
to come into contact. After a contact evolutionary phase  \citep{and+06},
the binary would merge to form a BSS accompanied by a main sequence
star in a very wide orbit. The contact phase may be rapid for 
low mass ratio binaries ($q<=0.6$), but could extend to a few $10^8-10^9$ yrs 
for binaries with higher mass ratios, before the final merger and the
formation of a BSS (\cite{nel+01}).
 
 In this figure we have used the equations
and parameters for KCTF found in \citet{2007FT} and the magnetic
braking prescription of \citet[\S~4.4]{egg06book}, in which angular
momentum is extracted from the spin $\Omega_{i}$ of each star $i=1,2$
at a rate: \begin{equation}
\frac{d\Omega_{i}}{dt}=-|\dot{M}_{i}|(\frac{2}{3}R_{i}^{2}+R_{A,i}^{2})\Omega_{i},\end{equation}
 where $\dot{M}_{i}=10^{-12}M_{\odot}$~yr$^{-1}$ are the mass-loss
rates from each component, $R_{i}=R_{\odot}$ are the stellar radii,
and $R_{A,i}=10R_{\odot}$ are the Alfv{é}n radii. This expression
corresponds to a saturated magnetic field \citep{and+06}, which is
valid for quick rotation. In reality, the stars lose a small amount
of mass which directly carries away a small amount of \emph{orbital}
angular momentum, but we have not explicitly computed those two effects.

\section{ Triple-formed blue stragglers interacting with cluster members }

\label{cluster BSS}Before discussing the implications of the triple
BSS progenitor model and its predictions for the BSSs properties,
we need to insure that such a scenario is viable for OCs or GCs where
the triple's evolution could be effected by encounters with other
stars. One should also note that close binaries may also be formed
through tidal capture encounters in the densest regions of GCs, i.e.,
in some cases they may not be formed from triples.

The studies of KCTF evolution of triples have usually dealt with isolated
triples. Triples evolving in clusters may be influenced by encounters
with other stars in the cluster \citep{2001AM,iva+08}. Several scenarios
are then possible: (1) an encounter destroys the triple before an
inner close binary forms; (2) an encounter occurs before an inner
close binary forms, but only perturbs the triple and does not destroy
it; (3) KCTF evolution produces an inner close binary before an encounter
occurs, and then an encounter destroys the outer binary of the triple,
leaving a close binary; (4) KCTF evolution produces an inner close
binary before an encounter occurs, and then either the triple evolves
as if it was isolated or it is perturbed but not destroyed.

In the first scenario the fraction of BSS progenitors is reduced relative
to their fraction in a similar population of isolated triples, since
in this case the destroyed triples could not form close inner binaries.

The second scenario, however, suggests an interesting possibility.
In this case perturbed triples change their orbital paramters in a
chaotic way due to the encounter (see, e.g., \citealp{heg75,hut83,bin+87}
for the behavior in binary encounters), but are not destroyed. These
could be thought of as new triples. These new triples are now subjected
to the same possible scenarios as the primordial triples, i.e., some
of them could now form close inner binaries, while others are perturbed
or destroyed beforehand.

The third scenario would produce close inner binary progenitors of
BSSs. Such binaries are very hard (i.e., have orbital energy $E\ll m\sigma^{2}$;
where $\sigma$ is the velocity dispersion in the cluster and $m$
is the mass of the binary system; \citealp{heg75}), and are not likely
to suffer from further perturbations, and therefore effectively evolve
in isolation, and contribute to the fraction of close binaries and
the fraction of BSSs in the cluster. Though hard, these binaries did
not contribute their orbital energy to the energy budget of the cluster
to affect its dynamical evolution as a whole; rather, the energy was
deposited as tidal heat and radiated away, as for tidally captured
binaries. Binaries formed in this way are expected to be observed
as close binaries or even contact binaries, without a triple companion
and therefore reduce the fraction of such binaries with tertiary companions
(which is close to unity for close binaries in the field; \citealp{2006T}).
BSSs formed in such binaries are likely to do so through coalescence
and are therefore likely to be observed as single BSSs. Nevertheless,
in some cases they may be observed as very close binaries during a
mass transfer epoch before coalescence. The observable predictions
from such a scenario may be difficult to disentangle from the case
of tidally formed binaries. However, tidally captured binaries are
expected to form only in the densest regions of cluster cores, whereas
the triple scenario could produce such binaries even in OCs or the
outskirts of GCs, since triples may suffer encounters in these regions,
but the rates of tidal captures are negligibly small.

The fourth scenario is maybe the most intriguing in terms of the observable
implications. In this case KCTF evolution produces a close inner binary
in a system that survives as a triple in the cluster. The close inner
binaries may form a BSS, either by angular momentum loss through magnetized
winds, or by the primary evolving to its Roche lobe, prompting mass
transfer and coalescence. Such systems would therefore be observable
as long period BSS binaries (or a triple if the inner binary has transferred
mass but not yet coalesced), with period-eccentricity distribution
similar to that of the outer binaries in triples. If the triple is
later perturbed, a companion star is still predicted by the model,
but the orbital configuration of the outer binary is not predictable
by KCTF alone (as in \citealt{2007FT}).

\begin{figure}
\begin{tabular}{c}
\includegraphics[clip,width=0.9\columnwidth]{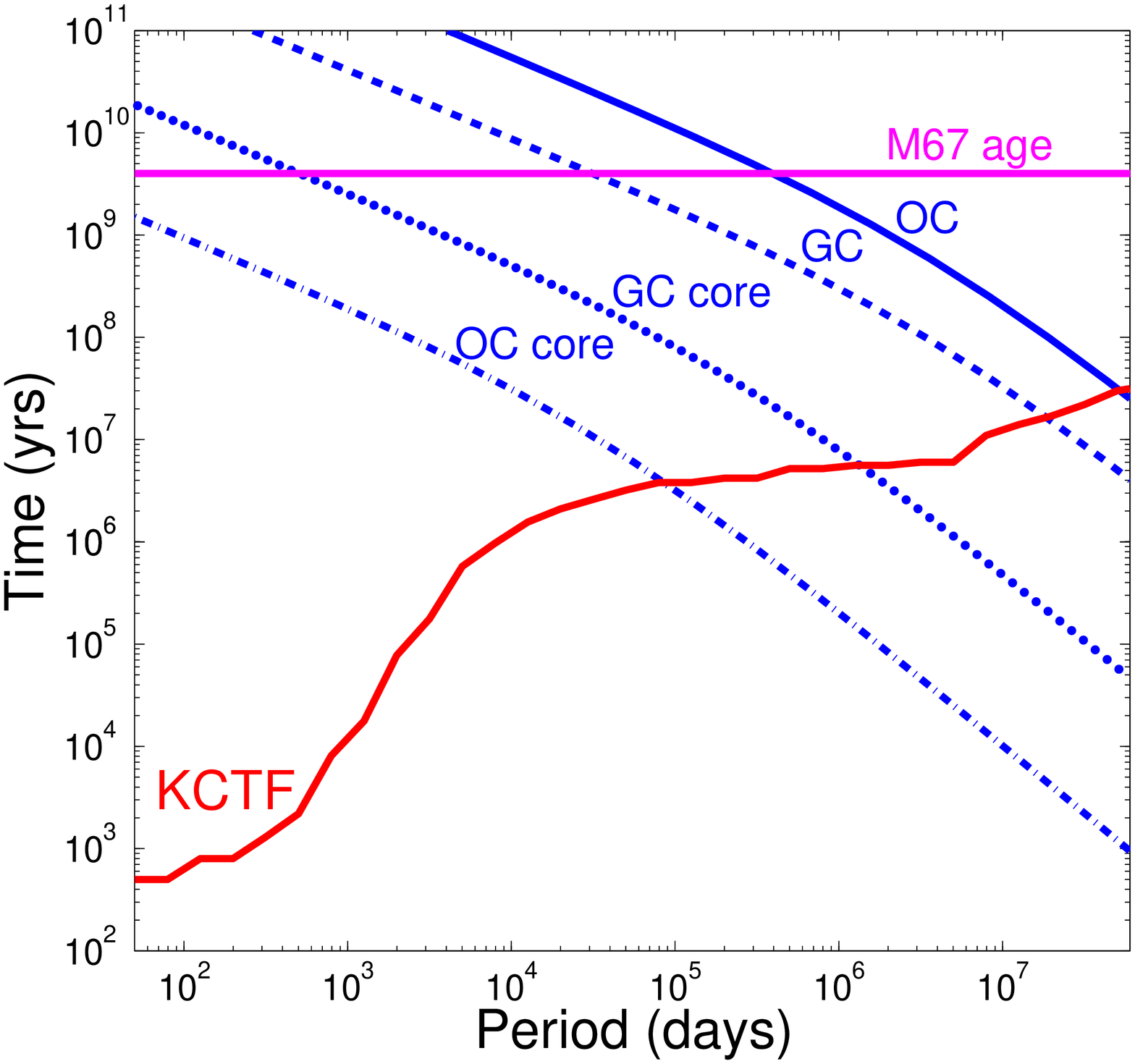}\tabularnewline
\end{tabular}

\caption{\label{f:encounter_time}Timescales for encounters and Kozai cycles
with tidal friction (KCTF) evolution of triples in clusters. The typical
encounter time of triples of different outer periods is shown for
different environments; OCs (upper solid), OC cores (dashed line),
GCs (dotted line), GCs cores (dash-dotted line). The typical KCTF
evolution time (see text) of triples with a range of outer period
is also shown (lower solid line). The typical age of GCs and the age
the old OC M67 are denoted by vertical lines).}

\end{figure}

In order to estimate the importance of the different scenarios we
can compare the typical timescales of the isolated KCTF evolution
and the typical time scales between encounters. In order to do so,
we used the methods described in detail by \citet{2007FT} to evolve
a large population of triple systems in isolation. We ran a Monte-Carlo
simulation of the evolution of primordial triples drawn from appropriate
distributions (as described in \citealt{2007FT}, where all inner
binaries were assumed to have initial periods of $P>6$ days) and
the triples were checked for stability using the criterion of \citet{mar+01}.
About $40$\% of the selected triples failed to fulfill the condition.
A total of $5\times10^{4}$ stable systems were then integrated in
time up to $10$ Gyrs, while neglecting stellar evolution or angular
momentum loss through magnetized stellar winds. In fig. \ref{f:encounter_time}
we show the typical timescale for the KCTF mechanism to form close
inner binaries from triples of different periods. This timescale is
defined as the median time close inner binaries take to become {}``close''
(defined as $P_{in}<6$ days, where magnetic braking may become important).
The feature apparent at $P_{out}=10^{3}-10^{4}$~d is a combination
of (1) the assumed eccentricity distribution of outer binaries switches
from a Raleigh distribution (moderate eccentricities) to a thermal
distribution (generally large eccentricities) there \citep{duq+91,2007FT},
and (2) inner binaries with companions at $P<10^{3}$~d must be initially
rather close to satisfy dynamical stability anyway, so they do not
have far to travel before magnetic braking becomes important. Also
shown are the typical encounter time scales for triples at different
periods, calculated for several cases; typical conditions in GCs,
OCs, GC cores, or OCs cores. 
The encounter time scale is given by (e.g., \citealt{iva+08}) \begin{eqnarray}
t_{enc}=8.5 & \times & 10^{12}yr\times P_{out}^{-4/3}M_{tri}^{-2/3}n_{5}^{-1}\sigma_{10}^{-1}\nonumber \\
 & \times & \left[1+913\frac{M_{tri}+\left\langle M\right\rangle }{2P_{out}^{2/3}M_{tri}^{1/3}\sigma_{10}^{2}}\right]^{-1}\end{eqnarray}
 where $P_{out}$ is the outer binary's period in days, $M_{tri}$
is the total triple mass in $M_{\odot}$, $\left\langle M\right\rangle $
is the mass of an average single star in $M_{\odot}$, $\sigma_{10}$
is the velocity dispersion $\sigma_{10}=\sigma/(10\, km\, s^{-1})$,
and $n_{5}$ is the stellar density in units of $10^{5}\, pc^{-3}$.
For simplicity we assumed all triples have equal masses of $M_{tri}=3$
with the average mass of stars $\left\langle M\right\rangle =1$.
This timescale is for encounters between a single star and a binary
(outer binary of a triple in our case). In binary-binary encounters
the cross section for the encounter is determined by the wider binary.
In the next section we compare the KCTF timescale to the encounter
timescale in a variety of environments, allowing us to establish predictions
for the observations. %

\section{Implications of the triple origin of BSSs in various environments}

\label{sub:BSSs-low density}

\subsection{Low density environments (OCs and GC outskirts)}

In relatively low density environments such as OCs or at the outskirts
of GCs, the timescales for close encounters is larger than the Hubble
time (and could be much larger than the typical age of OCs). In these
clusters the triples effectively evolve in isolation. The only caveat
is for very long outer period triples $(P_{out}\gtrsim5\times10^{4}-5\times10^{5}$
for the conditions in OCs). Such triples can form close inner binaries
in $\sim10^{7}$~yr that evolve into a BSS, but the outer binary
may still encounter other stars in the cluster later on and therefore
change its configuration. In particular, triples for which the KCTF
mechanism was inefficient may be perturbed into a different configuration
which is more favorable for KCTF evolution. Since the timescale for
the Kozai cycles, when those occur (i.e., $40^{\circ}\lesssim i\lesssim140^{\circ}$),
is usually much shorter than the time between dynamical encounters,
encounters are not likely to interfere with KCTF evolution and destroy
potential BSSs progenitors. Encounters can, however, contribute to
the formation of new BSSs progenitors. For example, if a triple is
initially coplanar it cannot evolve through KCTF evolution to form
a close inner binary. However, if an encounter excites its mutual
inclination sufficiently, then KCTF will rapidly operate before the
next encounter. If the companion stays bound, it has a better chance
of causing KCTF evolution. Therefore encounters that do not destroy
the triple tend to make KCTF evolution more likely (for a similar
discussion in the context of the KCTF mechanism in dense environments,
see \citealt{per+08}). Since tidally captured close binaries are
not expected to form in these low density regions, we can conclude
that the majority of close binaries should be part of triple systems,
and in particular, the majority of BSSs produced in close binaries
should be part of triple systems. Such a conclusion has many implications
for BSSs properties, some of which can be compared with currently
available observations to test the triple scenario, while others can
be checked by future observations. In the following we list these
implications and compare them with observations, when available. The
triple origin of BSSs has many implications and direct falsifiable
predictions.

\subsubsection{BSS fraction}

The BSSs formed in the triple scenario mostly form through the merger
of the inner close binary in a triple. As discussed above it is likely
that nearly all close binaries (defined as $P<6$ days) are formed
in triples. The BSS fractions we expect to be produced are therefore
similar to those predicted in the close binaries merger scenario,
where the only difference is that we expect the produced BSSs to have
companions (see below). \citet{and+06} studied the BSS fractions
from mergers of close binaries. Their results for the BSS fractions,
applicable also to the triple scenario, are that the expected BSS
fractions are consistent with observations. Note, however, that such
results require a relatively high fraction of close binaries to be
assumed. Such caveat applies in general to other theoretical predictions
in the literature (e.g. \citealp{hur+05} that also assumed high fraction
of close binaries). In the triple scenario such high fraction of close
binaries is the result of KCTF evolution, and need not be primordial.
This is important since it is not clear how can binaries with $\lesssim6$
days period form primordially, as the protostars of the binaries components
are of comparable size to the size of such short binaries \citep{2007FT}.
Binary distributions containing large fractions of close binaries
are indeed observed in very young clusters and are therefore reasonable
assumptions (see extended discussions in \citealp{hur+05} and \citealp{and+06}).
Note however, that the field distribution of binaries in the binaries
sample of \citet{duq+91} show smaller fractions of close binaries
than those observed in very young clusters, which is perhaps the signature
of their merger.

\subsubsection{BSS binaries fraction}

BSSs formed following the merger of the inner binary of a triple should
still have a companion following the merger. Although such binaries
could later be destroyed through encounters with other stars, this
is likely to happen mainly for very large period binaries (soft binaries).
The binary fraction of BSSs should therefore be higher than the overall
binary fraction in the environment where they are observed. Current
observations of spectroscopic BSS binaries \citep{gel+08,gel+09,lat07}
could only detect binaries with periods of $\lesssim3000$ days to
a good level of completeness, due to the finite duration of the surveys.
Longer period binaries would be observed as single BSSs and could
therefore lower the \emph{observed} BSSs binary fraction. Nevertheless,
hard BSS binaries with an initial $P_{out}>3000$ days could be affected
by encounters and harden to become closer, and thus observable, binaries
(since BSSs are more massive than other cluster stars, they are more
likely to survive in binaries even in exchange encounters). In any
case, the BSS binary fraction is expected to much exceed the general
binary fraction (of all stars) in the cluster environment. Since the
hardening of binaries is dependent on the age of the cluster and its
density, we may expect the (detection limited) \emph{observations}
of BSS binary fraction to show higher fractions in older and/or denser
environments, as long as the period detection limit is smaller than
the expected final hardening period \citep{hil84}. For example, for
the typical conditions in NGC 188, hard binaries should harden down
to periods of $\lesssim10^{4}$~days (see Eq. (5) in \citealt{hil84}),
this period is close to the detection limit of the binary periods,
and therefore most of the BSS binaries should be observed as such
and the observed BSS binary fraction is expected to be high ($>0.5$),
as indeed confirmed by observations (e.g. a binary fraction $>76\,\%$
for BSS binaries in NGC 188 \citealp{gel+08}). M67 is a younger and
less dense cluster and therefore the expected BSS binary fraction
is expected to be lower than that in NGC 188, but still higher than
the general binary fraction in the cluster.

We expect some specific correlations between BSSs fraction and cluster
properties in the triple KCTF scenario. The primordial triple fraction
is expected to be directly related to the binary fraction and we should
therefore expect a correlation between the binary fraction in clusters
and the BSSs fraction. Since close encounters do not have important
effects on the BSSs formation in the triple KCTF scenario, the BSSs
fraction should be related to the unperturbed evolution of triples,
similar to scenarios of unperturbed evolution of binary stars, which
is consistent with recent analysis of the correlations between cluster
properties and BSS fractions\citep{dav+04,sol+08,kni+09}.

\subsubsection{The period-eccentricity distribution of BSS binaries}

\label{sec:pedist}

\begin{figure*}
\begin{tabular}{c}
\includegraphics[clip,scale=0.35]{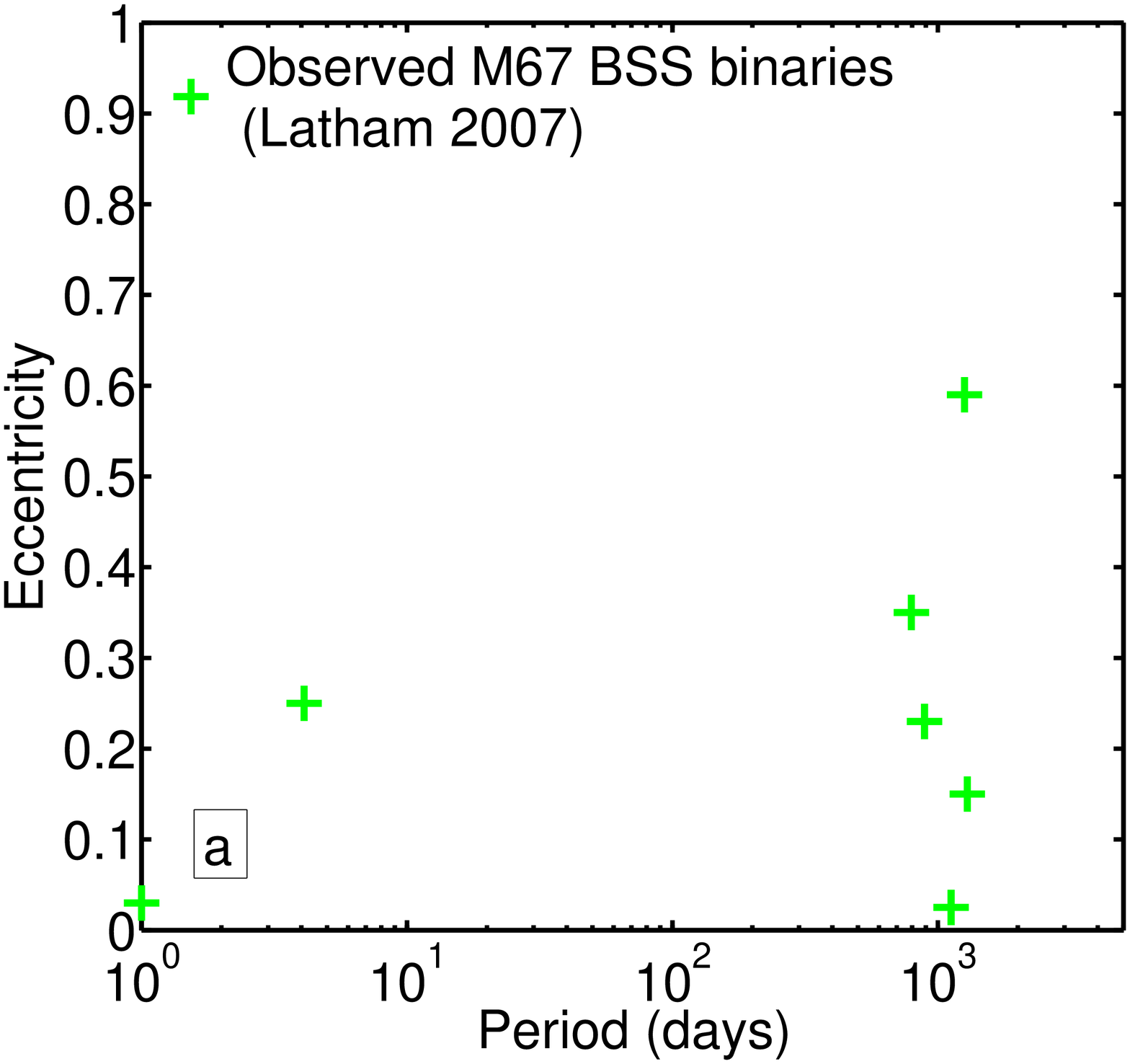}\includegraphics[scale=0.35]{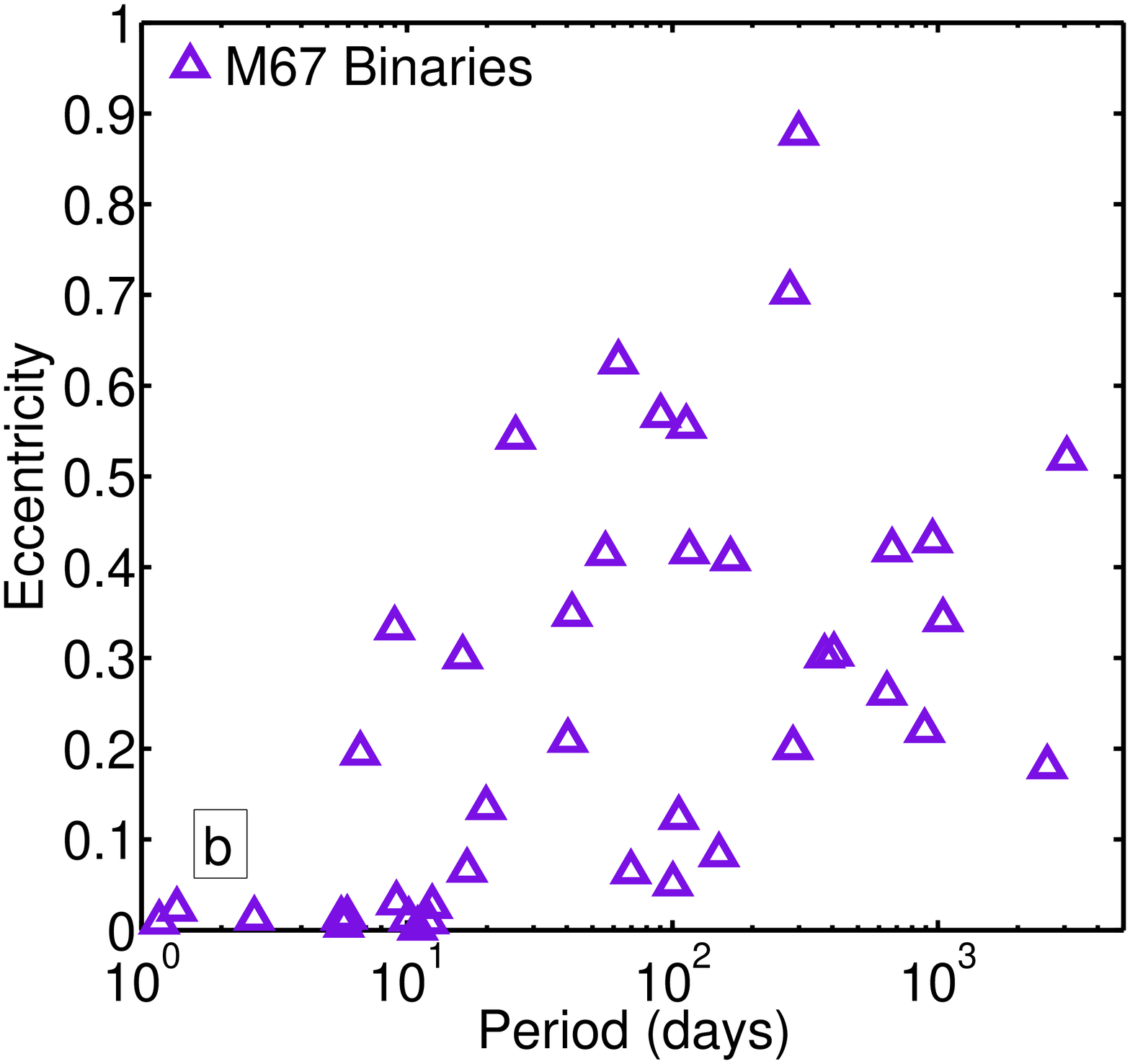}\tabularnewline
\end{tabular}

\includegraphics[scale=0.35]{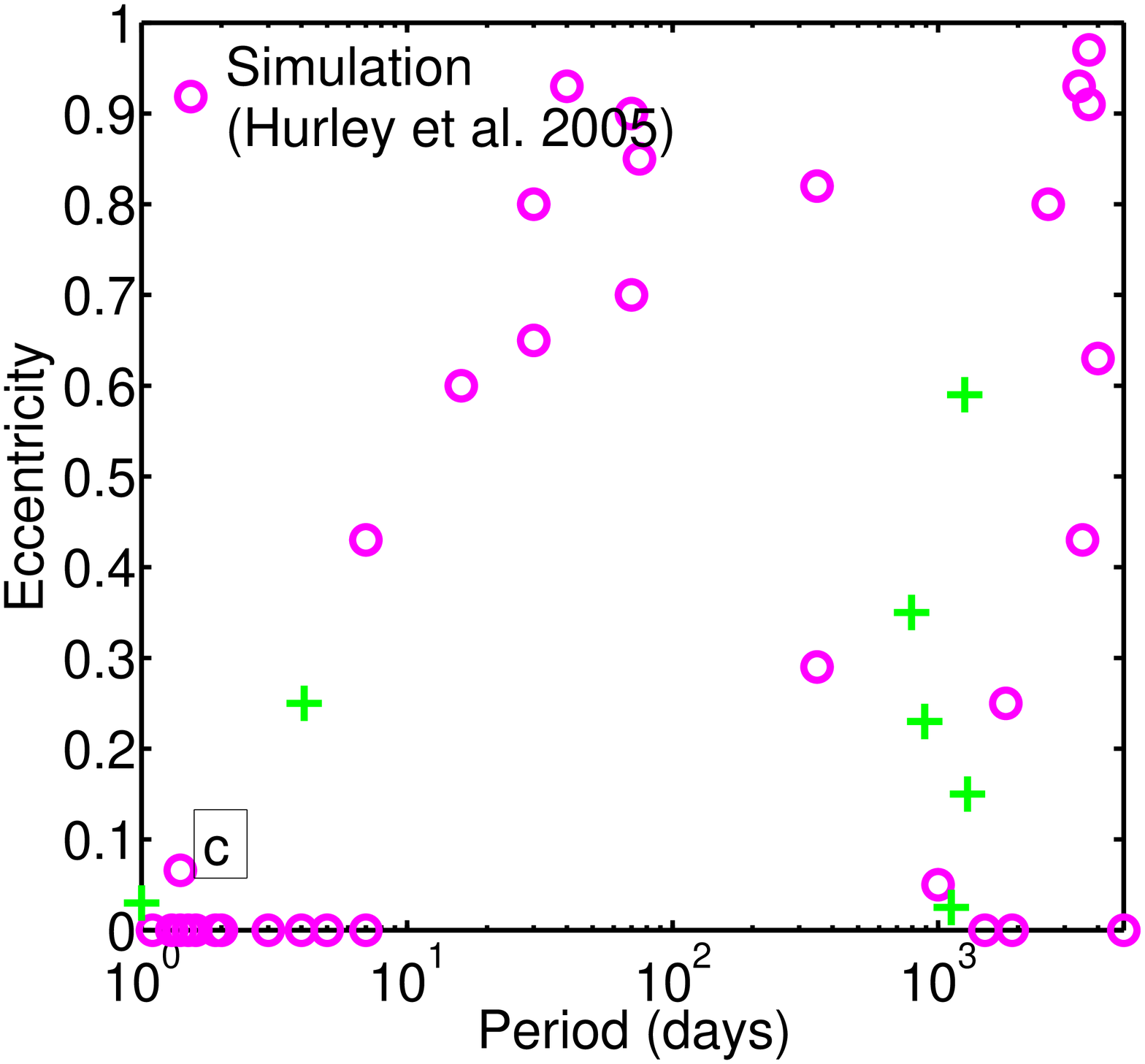}\includegraphics[scale=0.35]{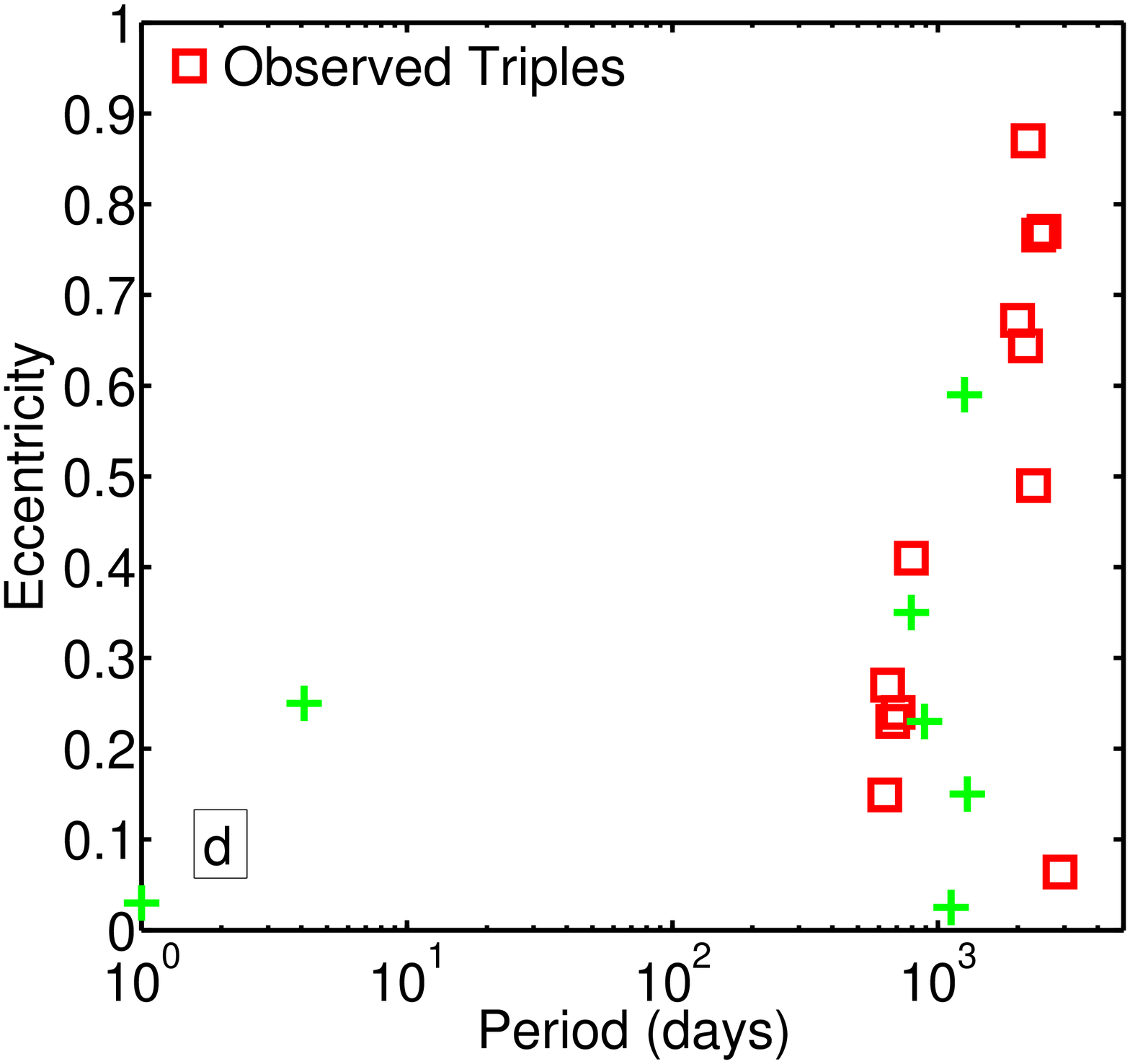}

\caption{\label{f:period-ecc}The period-eccentricity distribution of observed
BSS binaries. (a) The period-eccentricity distribution of BSS binaries
in M67 ($+$; \citealt{lat07}), note that the larger sample of BSS
binaries in NGC 188 (not shown; \citealt{gel+09}) show similar behavior
to that of M67 BSS binaries . (b) The period-eccentricity distribution
of regular (non-BSS) binaries observed in M67 ($\triangle$; \citealp{lat07}).
(c) The period-eccentricity distribution of BSS binaries produced
in N-body simulations of M67 ($\circ$; \citealp{hur+05}) compared
with the observed BSS binaries. (d) The period-eccentricity distribution
of outer binaries in triple systems with close inner binaries ($\square$;
taken from the multiple stars catalog \citep{tok97}) compared with
the observed BSS binaries. Only binaries with periods shorter than
$3000$ days are shown (approximately the radial-velocity detection
limit for the periods of the BSS binaries in the clusters). The good
agreement between the distribution of outer binaries of triples to
that of BSS binaries is evident. The comparison of BSS binaries to
the other distributions (regular binaries in the same cluster or simulated
BSS binaries in \citealt{hur+05} simulation) is poor. }

\end{figure*}

The triple origin scenario implies that the period and eccentricity
distribution of the BSS binaries should be the same as that of the
outer binaries of triple systems (observed in the field) that have
close inner binaries. In fig. \ref{f:period-ecc}d we show such a
sample of such triples. The inner binaries in such field triples with
$P_{in}<6$ days should be directly comparable with BSSs progenitors,
although we may miss some of the closest inner binaries in the field
which may have already merged via magnetic winds. Systems seen as
triples with close inner binaries now are likely to become BSS in
binaries later. We therefore also show the sample of triples with
larger inner periods ($P_{in}\lesssim10$~days), which is also likely
to be comparable, and could add to our statistics. We choose an upper
cutoff of $3000$ days, comparable to the detection limit of binaries
in the observations of M67 and NGC 188. All the triples were chosen
from the multiple stars catalog. Only low mass triples, i.e., triples
not containing stars with masses larger than $\sim3\, M_{\odot}$were
chosen. Higher mass stars would evolve off the MS on short timescales
and not contribute to the BSSs in older clusters such as M67 and NGC
188.

The period-eccentricity distribution of the outer binaries in the
triples discussed above is in good agreement with the observed BSS
binaries distribution in the OC M67 (see figure \ref{f:period-ecc}d)
and NGC 188 (full data will be available in \citealp{gel+09}). Although
the triples sample is not large, one can still observe some very unique
characteristics that would be expected for BSS binaries in the triple
origin scenario; note that the general binary distribution in fig.
\ref{f:period-ecc}b clearly shows a different behavior than that
of the BSS binaries in fig. \ref{f:period-ecc}a. In particular, we
expect BSS binaries from the triple scenario to usually have large
periods, typically with $P\gtrsim700$ days (also true for the larger
NGC 188 sample, where 12 out 15 BSS binaries have periods longer than
this cut-off\citealp{gel+09}) %
\footnote{Interestingly, higher mass triples do show outer binary periods much
shorter that $\sim700$ days, which suggest very different characteristics
of massive versus low mass triples (see also \citealt[\S~3.2.1]{per08a},
in this respect), which may suggest a very different BSS binary period
distribution in very young OCs.%
}. Such a lower cut-off for the period of the outer binaries in triples
could be the result of their formation process. Closer triples may
have formed with correlated angular momentum. In such a case the relative
inclination between the inner and outer binaries in the triple may
be low or even close to zero, which will quench KCTF evolution. These
triples could not then produce close inner binaries. Note also that
the stability of the triples may also play a role in biasing against
the formation of close triples \citep{2006T}.  We also note that
observed BSS binaries with $P<10$ days are more likely to be the
inner binaries of triples in which the BSSs were rejuvenated through
mass transfer and not through a full merger (similar to the close
BSS binaries produced in \citealt{hur+05} simulations). Since the
rejuvenated BSSs have only accreted some of their companions' mass,
such close BSS binaries are likely to have lower masses (and be fainter)
than the typical BSSs observed in the cluster. Even within the triple
scenario, then, it is no surprise that some short-period BSS binaries
do not match the $P$-$e$ diagram of \emph{outer} binaries; the scenario
predicts that a third, yet unseen star orbits each of these binaries.

Regarding the eccentricity distribution of BSSs produced by the triple
scenario, we note that the eccentricities of outer binaries are consistent
with the distributions of regular binaries in the field \citep{1991DM}:
generally low ($\lesssim0.4$) eccentricities for period of $P\lesssim1000$
days and a wider distribution up to high eccentricities for periods
of $P>1000$ days (again also consistent with the larger sample in
\citealp{gel+09}). For the triple to be initially dynamically stable,
the periastron of the outer binary must not pass too close to the
inner binary, and this constraint translates to an upper limit on
the eccentricity of the outer binary (which becomes the BSS binary),
although for individual systems this limit is difficult to evaluate
as the orbit of the original inner binary is not known.

The unique properties of the outer binaries in triples which are consistent
with the behavior of BSS binaries is difficult to reproduce by either
the collisional or the binary mass transfer scenarios for BSSs formation,
even when combined together with the full dynamics of the system (see
fig. \ref{f:period-ecc}c; high eccentricities are mainly due to exchange
encounters). The observations of BSS binaries therefore serve as a
good discriminator between different formation scenarios, and current
observations clearly favor the triple scenario as the major formation
route of BSSs.

\subsubsection{The companions of blue straggler }

The binary companions of long period BSSs could be MS stars in the
triple scenario (see also \citet{2006EKE,egg+08} for a related discussion,
and possible observations of such cases), which could not be the case
for long period BSS binaries produced following the post MS evolution
of their companion, that are expected to be white dwarfs (WDs) at
this stage. Another interesting discrimination method in this respect
might be the comparison to CH stars found in clusters (which have
low eccentricities and long periods, consistent with post-MS evolution
in binaries; e.g., \citealp{mcc97}). If the triple KCTF scenario
is the dominant mechanism for BSSs formation, one would expect very
different distribution of CH binary stars and BSSs binaries, but very
similar distribution if the mass transfer scenario is the dominant
mechanism, and the former is apparently the case.

Since higher multiplicity systems are abundant (e.g., quadruples are
1/3 as abundant as triples), one may find a similar fraction of the
BSSs binaries to be in triple (or higher multiplicity) systems. In
particular, two binaries in a double-double quadruple system (see,
e.g., the doubly eclipsing lightcurves of \citealp{2007PS}) could
produce a long period binary system containing two BSSs. This could
happen if both inner binaries merged. Similarly a triple (or quadruple)
system containing two BSSs could form in this way if one (or both)
of the inner binaries transferred mass, but has not merged, yielding
BSSs in the inner binaries. We therefore expect multiple systems containing
more than one BSS to be observed\footnote{Interestingly, a triple containing 
two BSSs, one with a close companion are observed in M67 
\cite{van+01,san+03}. Such system could be formed, in principle, 
through the quadruple evolution we discuss.  
However, the large masses of s1082 components inferred from the orbital
solution  (Sandquist et al. 2003; but note the discrepancy with the position
in the color magnitude diagram), would require such system to be much younger
(age of ~1-1.5 Gyrs) than the age of M67 (4 Gyrs).}.
.

\subsubsection{The masses of blue stragglers and their location in color magnitude
diagrams}

\label{sub:bss-masses}

Mass transfer in long period binaries ($P>700$ days, such as the
BSS binaries observed) is highly inefficient. The total mass transferred
from the post MS companion to the formed BSS is likely to be small
($<0.3\, M_{\odot}$; and typically even lower) in this case, producing
BSSs with masses not much larger than the turn off mass of the cluster
(i.e., $M_{BSS}\lesssim M_{turnoff}+0.3\, M_{\odot}$). The triple
scenario can produce a much more massive BSS, as it is the sum of
the inner components of the triple ($M_{BSS}\lesssim2M_{turnoff}$).
Such a higher mass would therefore discriminate them from those formed
through mass transfer. The latter binaries are expected to be composed
of a low mass (fainter) BSS with a WD companion at an intermediate
period of a few hundred days as observed for field BSS binaries (see
\citealp{pre+00} and section \ref{sec:field BSS}). Such BSSs would
be located close to the turn-off mass in the color magnitude diagram
of a given cluster, whereas the BSSs from the triple scenario can
be distributed much further, typically far from the turn off point
of the cluster. Note that most BSSs in both NGC 188 and M67 are far
from the turn-off point in the color magnitude diagrams of the clusters
\citep{gel+08,liu+08}, consistent with the predictions of the triples
scenario and at odds with the binary mass transfer origin. 

In case a close BSS binary did not fully merge, i.e., a close companion
could still be detected, the BSS product should be be less massive,
than the product of a full merger, as mentioned before, 
and would be observed closer to the turn off point
of the cluster than a merged BSS. The binary companion of such BSSs is likely 
to be an evolved star, and the system could be a close or even 
contact binary (possibly an Algol like system, e.g. \cite{jeo+06,kal+07}).
The frequency of such BSS Algols and other close BSS binaries, 
is still unknown, but is likely to be small (\cite{tia+06}).
In addition one should then expect to find a triple
companion, as is observed in close binary systems in the field (see
\citealp{sep+00} for possible observation of such systems).

In some cases the companion of the KCTF-formed BSS may evolve off
the MS during the lifetime of the BSS, in which case the BSS may accrete
some of its mass, and become a more massive BSS, possibly even extending
beyond twice the turnoff mass of the cluster. This could possibly
explain the existence of some over-massive BSSs observed in clusters.
In such cases we might expect the binary companion to be a WD, probably
on a rather circular orbit.

\subsubsection{Spin-orbit correlation in BSS binaries}

In their theoretical analysis, \citet{2007FT} found that the distribution
of the relative inclination between the inner and outer binary orbits
of KCTF triples should be in the range $30^{\circ}-150^{\circ}$,
with peaks at $40^{\circ}$ and $140^{\circ}$. This distribution
is therefore expected for BSS triples with non-fully merged inner
close binaries. If the inner binaries did merge, it is likely that
the angular momentum of the pre-merged inner binary would leave its
signature on the spin of the merged BSS. It is therefore possible
that the typical relative inclination distribution found by \citeauthor{2007FT}
could still be detectable in relative inclination between the \emph{spin}
of the BSS and the binary orbit (i.e., the \emph{obliquity} of the
BSS), which is observationally accessible (although currently only
through an overall statistical analysis; e.g., \citealt{1994H}).

\subsubsection{Radial distribution of BSSs in clusters}

Since our triple scenario suggests the same progenitors for BSSs and
close binaries, we expect the radial distribution ( from the cluster
center) of BSSs in clusters to be similar to that of close binaries.
This prediction may seem natural also in the scenario of BSS production
from primordial close binaries not evolved in triples, however there
are few differences between the two predictions. In principle, triple
systems are likely to be more massive then binary systems, and therefore
be somewhat more mass segregated in clusters, and possibly have different
formation efficiency in different parts of a cluster, than that of
binaries. These could possibly make apriori differences in the radial
distribution of BSSs compared with regular binaries, although it is
not clear whether enough statistics exist for making such a signature
significant. In addition in the triple scenario we would also expect
BSS binaries with long periods (i.e., with the inner binaries fully
merged, see \S~\ref{sec:pedist}) to have a similar radial distribution
to that of regular (non-BSS) close binaries in the cluster. If short
and long period binaries have different radial distributions, the
BSSs distribution should also reflect this. 

Possible differences in the radial distribution may also serve as
good discriminators between the triple scenario and the combined effects
of all other suggested scenarios such as studied in N-body simulations
(\citealp{hur+05}), since the latter suggest that BSSs and especially
BSS binaries form only in the inner regions of the cluster whereas
the triple scenario could also form BSSs and BSS binaries in the outskirts
of clusters.%
\footnote{Note that observation of a bimodal radial distribution of close binaries
in a cluster would be highly interesting, since such a distribution
is usually thought to be quite unique for BSSs. Following this prediction
we searched the literature for any evidence of such bimodal radial
distribution of close binaries in clusters, and indeed found two examples
for such a distribution in the clusters $\omega$~Centauri and Tuc
47 \citep[this is by no means a complete list, and many others may exist]{wel+04,wel+07}.
The analysis of this phenomena is, however, beyond the scope of this
paper and will be discussed elsewhere. %
}

\subsubsection{A general relation between BSSs and close binaries }

A more general prediction of the triple KCTF scenario is the close
relation between close binaries and the BSSs population. This relation
suggests that the predictions and implications described in the previous
points regarding BSSs should be applicable also to the general population
of close binaries in clusters (e.g., eclipsing binaries, contact binaries
W UMa binaries, etc.) which may also be expected to have third companions
in many cases. We note however that some close binaries populations
such as X-ray binaries and cataclysmic variables were likely to form
in a different route. In such populations the close binaries form
only after the main sequence evolution of the compact object progenitor,
and the binary separation must have been large during this time (e.g.
\citealp{rit+08} and references therein), i.e. excluding the possibility
of KCTF evolution in which the binaries must have very small pericenter
distance during their evolution.

\subsection{High density environments (GCs cores)}

As can be seen in fig. \ref{f:encounter_time} the timescales for
close encounters with triples is larger than the typical KCTF time
even in the dense environments of GCs, and therefore close binaries
could form through KCTF even in such environments. Therefore the primordial
triples (and dynamically formed triples; see \citealp{iva08,iva+08,2008T}),
should be taken into account when studying BSSs formation scenarios
in GCs. Nevertheless, given the high encounter rates in GCs, even
close KCTF-formed binaries may be involved in close encounters which
could destroy them, change their configuration or cause collisions,
and possibly forming BSSs. The outer binaries in such triples are
even more likely to be involved in several encounters at some stage
during the GC evolution, which may even disrupt the triple, if the
outer binary's periastron sinks too close to the inner binary (see,
e.g., \citealp{heg75,hil84,aar04}). Given that the BSSs that do form
would have a long period binary companion, they are most likely to
be involved in many encounters, following which they are still expected
to be in relatively wide binaries (for the hard triples), even in
the case of an exchange. According to \citet{2008T}, $\sim1\%$ of
binaries have tertiary companions at any given time by dynamical formation,
but this population continually evolves, so up to $\sim10\%$ of binaries
have a chance to evolve through KCTF. The complex dynamics of high
density environments (see, e.g., \citealp{del+97,aar04}) and the
need to include stellar evolution in old system such as GCs makes
it difficult to make clear predictions for the BSSs population without
more elaborate dynamical analysis and/or N-body simulations; it is
beyond the scope of this work.

In addition, given the old ages of GCs, BSSs formed from primordial
triples in the triple scenario may already evolve off the MS in these
clusters (see also next section), and possibly only the dynamically
formed triples could then directly form BSSs at later stages of the
cluster evolution (see also \citealp{iva08,iva+08}). In this case
BSS binaries observed in GCs would have a different distribution than
that of BSS binaries in lower density environments, with likely more
eccentric and shorter period orbits.


\subsection{Halo environment }

Now we discuss BSS formation, including the triple scenario, within
the old populations of the halo of the Milky Way and of early type
galaxies, which have properties similar to old OCs and the outskirts
of GCs.

\subsubsection{Field BSSs}

\label{sec:field BSS}

In the galactic halo the stellar population is expected to be very
old, so apparently younger stars are conspicuously blue; these are
called Field BSSs {\citep{pre+00,car+01,car+05}}. Such a non-cluster
stellar population is expected to evolve in isolation, and therefore
field BSSs are not expected to form through collisions. Field BSSs
could be explained by the binary stellar evolution of isolated binaries
with periods of at most a few thousands days which evolve through
mass transfer to form binaries with typical periods of a few hundred
days ($100-800$) \citep{mcc64}. In such binaries one of the components
would evolve and expand, leading to a mass transfer to its companion
which could become a BSS. Such stellar evolution would usually lead
to circularization the binary orbit and its shrinkage and leave a
WD companion to the BSS. The field BSSs identified by \citet{pre+00}
and \citet{car+01} are single-lined spectroscopic binaries, so the
companions are consistent with being WDs, and the orbital period distribution
is consistent with that expected from mass transfer in evolved binaries.
One may ask whether triple systems could also contribute to formation
of field BSSs.

When it operates, the KCTF mechanism leads to a rapid formation ($\lesssim10^{7}$yr)
of a close inner binary. Stellar evolution of such a close binary
is likely followed by its merger%
\footnote{We do caution, however, that for wide systems for which the TF timescale
is too long initially, stellar evolution can play an important role
in the KCTF evolution, leading to a complicated interplay of mass
transfer and eccentricity driving. The analysis of such combined KCTF
and stellar evolution processes, however, is beyond the scope of this
work. See \citet{ibe+99} for foundational considerations on this
subject.%
}. A merger product is usually much more massive than a star which
accreted some mass from its companion (see \S\ref{sub:bss-masses}).
Therefore such a rejuvenated star is very likely to evolve off the
MS relatively early, and not be currently observed as a field BSS.
When evolved such a star may transfer mass to its long period companion
(originally the star in the outer period of the triple), forming a
low mass BSS (i.e at $M_{BSS}\lesssim M_{turnoff}+0.3\, M_{\odot}$).
Such a scenario may still show a weak signature on the BSS binary.

Since the outer period of triples is usually larger than $\sim700$
days (see \S\ref{sec:pedist}), field BSS binaries from triples may
have larger periods, on average (although these may be somewhat shortened
during the mass transfer evolution). The more massive BSSs may also
produce more massive WDs, on average. Combined together we might observe
wider period field BSS binaries to have more massive WD companions
(see \citealp{bon+01} for a related scenario). The most massive (and
luminous) field BSSs could be the product of a full merger of the
inner binary in triples with very long KCTF timescales. Observation
of a field BSS binary with a main sequence companion could serve as
a strong evidence for a triple origin in this non-collisional environment,
where an exchange scenario for the MS star origin is not possible.
Such a BSS is likely to be one of the brightest, most massive field
BSS. Nevertheless, the signature of triple formed field BSSs is not
strong, and it is possible that in most cases triple formed BSS binaries
would be indistinguishable from the other field BSS binaries produced
in binaries. We conclude that triples may contribute to the formation
of field BSS binaries, but their contribution is not likely to be
dominant, and would leave only a weak signature, unless very massive
BSSs are observed ($\geq2\, M_{\odot}$).

\subsubsection{Bright planetary nebulae}

\citet{cia+05} suggested that the progenitors of bright planetary
nebulae (PNe) observed in old stellar population are BSSs formed in
close binaries, since the progenitor mass of these bright PNe are
thought to be larger than $2\, M_{\odot}$. Such high mass is much
beyond the turn off mass of stars in old stellar populations, such
as in early-type galaxies and in galactic halos. Moreover, high mass
BSSs are likely to form only through mergers or strong mass transfer
(see \S\ref{sub:bss-masses}) which are only produced in close binaries
or in collisions in dense environments. Since in the triple origin
for BSSs we suggest such close binaries and their merged BSS product
are formed in triples, a straight-forward conclusion is that the progenitors
of bright PNe are also triple stars. We therefore predict that such
PNe may still have a long period binary companion after the inner
binary in the triple produced the BSS which evolved to become a PN.
The further evolution of the BSS to a PN could affect the binary orbit,
possibly circularizing it, due to low mass transfer from the evolved
BSS to its companion. Likewise, the presence of the binary companion
could affect the morphology of the nebula. The details of such evolutionary
process are beyond the scope of this work.

\section{Summary}

\label{s:Summary}In this paper we have studied the possible formation
scenario of BSSs in primordial and dynamically-formed triple systems
and its implications for the evolution and observations of BSSs. The
direct relation between triple stars and BSSs in this scenario suggests
a strong connection between BSSs properties and those of triples stars.
Many specific predictions for the BSSs populations are implied by
this relation and described mainly in section \ref{sub:BSSs-low density},
some of which are unique predictions that can discriminate it from
the two other BSSs formation scenarios: stellar collisions and mass
transfer in or merger of binaries. Possibly the strongest signature
expected from this scenario is the expected high binary fraction of
long period BSS binaries and their unique period-eccentricity distribution
with its strong bias towards long period orbits ($>700$ days). This
distribution is not likely to be produced by any other single scenario
for BSSs formation, and not even through their combined effect as
studied in N-body simulations with stellar evolution. We showed that
the recent observations of the BSS binaries population in the open
clusters M67 and NGC 188 (the only clusters for which we have a wealth
of data on BSSs binaries) could naturally be explained by the triple
scenario, where all other currently suggested scenarios for BSS formation
and evolution encounter major difficulties (see fig. \ref{f:period-ecc}).
The triple scenario is likely to play a more minor role in the formation
of field BSSs and in other low density \emph{old} stellar populations,
but could be important for the production of the most massive BSSs
in these environments. The brightest planetary nebulae observed could
be the product of such massive field BSSs, and may therefore have
long period binary companions as expected for BSSs in this scenario.
In the cores of globular clusters the interplay between triple evolution
and other dynamical effects may become more complex and both processes
are likely to play a role in the BSSs formation. However, in open
clusters the triple origin scenario is possibly the most dominant
mechanism for the formation of blue stragglers and currently the only
model explaining the BSS binary properties in these environments.

\acknowledgements{We would like to thank Bob Mathieu and Aaron Geller 
for helpful discussions and for supplying us with their pre-published 
data on BSS binaries in NGC 188. We thank Scott Tremaine and the referee
Peter Eggleton for helpful comments and discussions. 
HBP is supported by ISF grant 928/06.  gratefully acknowledges support from
a Michelson Fellowship, which is supported by the National Aeronautics
and Space Administration and administered by the Michelson Science
Center.}

\end{document}